\documentclass[conference]{IEEEtran}
\IEEEoverridecommandlockouts
\usepackage{amsmath,amssymb,amsfonts}
\usepackage{algorithmic}
\usepackage{graphicx}
\usepackage{textcomp}
\usepackage{xcolor}
\usepackage{subcaption}
\usepackage{microtype}
\usepackage{multicol}
\usepackage{bm}
\usepackage{booktabs}
\usepackage{cite}
\def\BibTeX{{\rm B\kern-.05em{\sc i\kern-.025em b}\kern-.08em
    T\kern-.1667em\lower.7ex\hbox{E}\kern-.125emX}}
\begin{document}

\title{Efficient Reprogramming of Memristive Crossbars for DNNs: Weight Sorting and Bit Stucking}

\author{\IEEEauthorblockN{Matheus Farias}
\IEEEauthorblockA{\textit{Harvard University} \\
matheusfarias@g.harvard.edu}
\and
\IEEEauthorblockN{H. T. Kung}
\IEEEauthorblockA{
\textit{Harvard University}\\
kung@harvard.edu}
}

\maketitle

\begin{abstract}
We introduce a novel approach to reduce the number of times required for reprogramming memristors on bit-sliced compute-in-memory crossbars for deep neural networks (DNNs). Our idea addresses the limited non-volatile memory endurance, which restrict the number of times they can be reprogrammed. 

To reduce reprogramming demands, we employ two techniques: (1) we organize weights into sorted sections to schedule reprogramming of similar crossbars, maximizing memristor state reuse, and (2) we reprogram only a fraction of randomly selected memristors in low-order columns, leveraging their bit-level distribution and recognizing their relatively small impact on model accuracy.

We evaluate our approach for state-of-the-art models on the ImageNet-1K
dataset. We demonstrate a substantial reduction in crossbar reprogramming by 3.7x for ResNet-50 and 21x for ViT-Base, while maintaining model accuracy within a 1\%
margin.

\end{abstract}


\section{Introduction}
Resistive compute-in-memory (CIM) crossbars offer a promising computing architecture for deep neural networks (DNNs), using analog accelerators to achieve fast, power-efficient matrix multiplication. Central to this architecture are memristors, which enable simultaneous data storage and processing, overcoming the data movement limitations of traditional von Neumann systems \cite{Mutlu2023, chakraborty2020, shafiee2016, donhee2022, abu2020, huo2022, Zou2021}.


DNN weights are encoded into memristor conductances by applying sequential voltage pulses, which significantly increases programming time. In conventional 1 transistor–1 memristor (1T1R) crossbar architectures, each memristor is programmed individually, with the transistor acting as a switch to isolate the target cell for precise conductance adjustment \cite{kuzum2012, alibart2012, eryilmaz2014, gao2015, gao20152, merced2016, zhang2021}.



Memristors, based on non-volatile memory technologies \cite{kugeler2008, abu2020, donhee2022}, offer high retention but have limited endurance, degrading after a finite number of reprogramming cycles. In CIM architectures, where write operations are more frequent, endurance is especially critical; for instance, a 1024x1024 resistive RAM crossbar used for 32-bit multiplication may fail within minutes\cite{endurance}. This issue is intensified when implementing large DNNs, as they require partitioned sections to fit onto relatively small, fixed-size crossbars. Consequently, each crossbar requires multiple reprogramming cycles to accommodate each partition, further straining endurance. We propose programming sorted weight sections to reduce reprogramming frequency, thereby extending the lifespan of memristor-based architectures without sacrificing efficiency.

The use of sorted sections minimizes memristor state mismatch during reprogramming by grouping crossbars of similar weights. To the best of our knowledge, this is the first implementation of sorted weight sectioning (SWS) \cite{farias2024} for crossbar reprogramming. Moreover, SWS enhances parallelization by distributing programming across multiple crossbars based on weight magnitude, thereby balancing the workload across threads and further optimizing reprogramming.

We further propose a reprogramming method based on the distribution of memristor states, leveraging the statistical properties of bitline-represented numbers. As bitwidth increases, the probability of the least significant bit being one approaches 50\%, reflecting an inherent randomness in lower-order bits. From a CIM perspective, this reveals a lack of correlation between the bell-shaped distribution of DNN \cite{han2015, fang2020, horton2022, tambe2020} and the uniform distribution of active memristors in low-order columns, which are farther from the input and contribute minimally to the overall weight magnitude. Recognizing their limited impact, we selectively reprogram only a small fraction of memristors in the lowest-order column, reducing the overall reprogramming workload without sacrificing performance.

\begin{figure*}
    \centering
    \includegraphics[width=.70\textwidth]{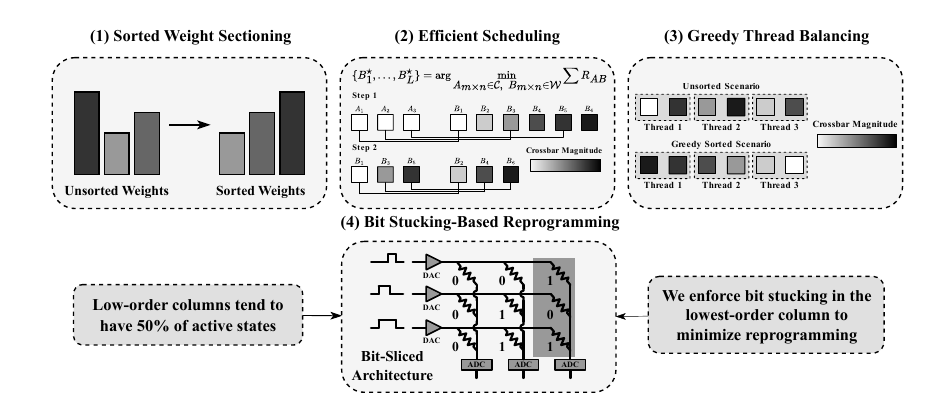}
    \caption{Summary of the approach. (1) We apply sorted weight sectioning, (2) efficient schedule crossbars with stride 1 to minimize transitional states, (3) assign similar crossbars to each thread in a greedy approach to minimize imbalance, (4) since low-order columns tend to have 50\% of active states, we enforce bit stucking in the lowest-order column to minimize reprogramming without severe DNN accuracy degradation.}
    \label{fig:summary}
\end{figure*}

See Figure \ref{fig:summary} for the summary of the approach. The main contributions of this paper are:
\begin{itemize}
    \item The use of sorted weight sectioning to (1) minimize reprogramming and (2) efficient scheduling of multiple crossbar programming without penalizing model accuracy
    \item A bit stucking procedure based on the uniform distribution of memristors in low-order columns to further reduce reprogramming with minimal DNN accuracy loss
\end{itemize}

\section{Background and Related Work}
\label{sec:back}



Past works addressed crossbar programming efficiency in terms of time and accuracy. Feedback strategies \cite{zhang2021, yu2023analog} uses multiple read/write operations to approximate the output of multi-level memristors. Variable amplitude pulses \cite{gao2015, corinto2018, kuzum2012} implement finer transitions between adjacent conductance levels. Emmanuelle et. al. \cite{merced2016} combines feedback and variable pulses and is considered as the state-of-the-art implementation of sequential programming. Recent papers addressed parallel memristor programming with sequential fine-tuning \cite{zhang2021} and neural networks assisting memristor programming \cite{yu2024}.

Our work targets bit-sliced crossbars, where each row corresponds to a single weight value and each column represents a power-of-two multiplier. For example, in a 128x128 crossbar with 16 power-of-two multipliers, we can represent $128/16=8$ weights per row. We label crossbars as 128x16 instead of 128x128 with 16 multipliers. Additionally, we do not implement feedback loops since bit-slicing already ensures accurate weight representation without the overhead of time-consuming read/write operations \cite{abu2020}. Furthermore, we opt for constant amplitude phases to avoid the increased power consumption associated with variable amplitude phases.

\section{Optimizing Reprogramming with Sorted Weight Sectioning}
\label{sec:sorted_sec}
In bit-sliced crossbars, reprogramming is restricted to memristors changing states. To minimize these transitions, a weight allocation method called sorted weight sectioning (SWS) is used, where weights are sorted by magnitude and grouped into sections to reduce the frequency of state changes. This sorting is done once offline, but inference requires a buffer to batch data and apply index matching for correct dot-product calculations. Building on SWS, this approach reduces reprogramming and enhances energy efficiency as discussed in \cite{farias2024}.


\subsection{Reprogramming Crossbars}
\label{sec:reprogram}
Consider we have two crossbars $A, B \in \mathbb{R}_{m\times n}$, with memristors $a_{ij}, b_{ij} \in \{0,1\}$ representing, respectively, inactive and active states, for $i\in \mathbb{R}_m$ and $j\in \mathbb{R}_n$. The reprogramming cost to convert from $A$ to $B$ is
\begin{equation}
    R_{AB} = \sum_{i,j}|a_{ij} - b_{ij}|.
\end{equation}

If $|a_{ij} - b_{ij}|=0$, the memristor $a_{ij}$ does not face a transitional state. On the other hand, if $|a_{ij} - b_{ij}|=1$, we switch states. 
Mathematically, we want to find $B^{\star}$ such that 
\begin{equation}
    B^{\star} = \arg \min_{B_{m\times n} \in \mathcal{W}}R_{AB},
\end{equation}
where $\mathcal{W}$ is the set of all weight matrix sections of size $m\times n$.

\begin{figure}
    \centering
    \includegraphics[width=\columnwidth]{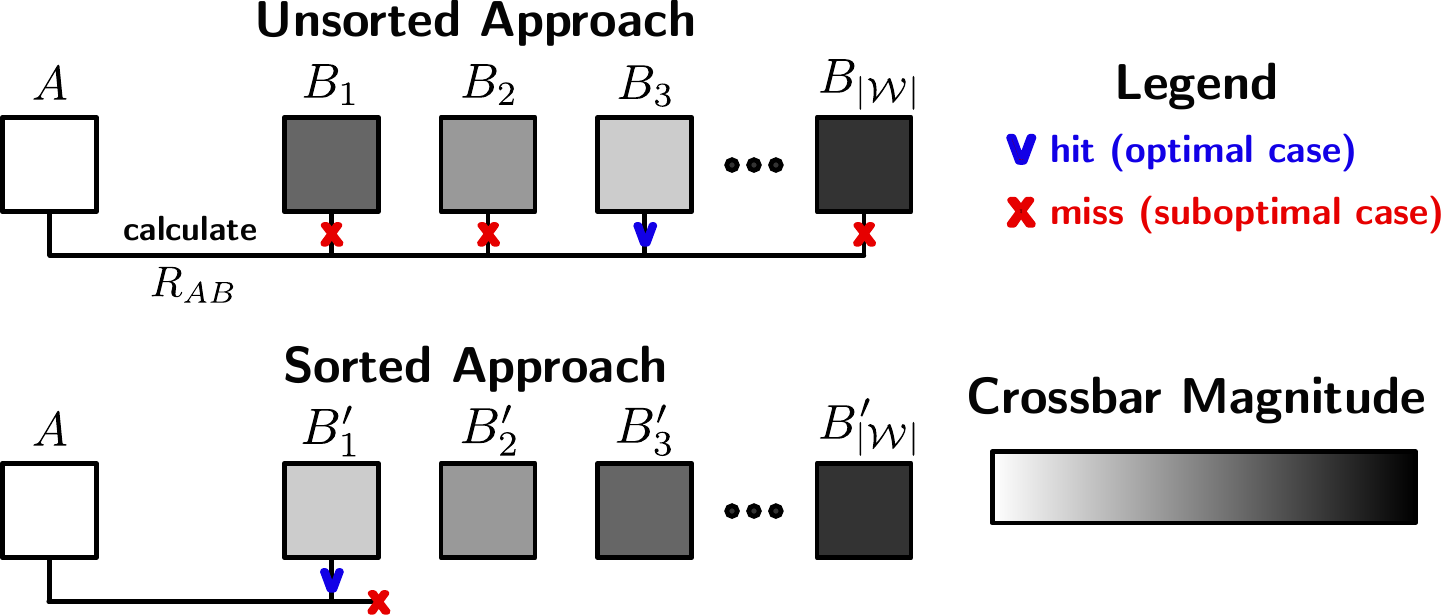}
    \caption{The unsorted approach calculates reprogramming costs for all sections, while the sorted approach picks the next crossbar to reduce state mismatches.}
    \label{fig:schedule}
\end{figure}

Without SWS, optimal scenario is achieved by iterating through all sections and calculating the cost to see which one provides the minimum effort. With SWS, we program the next section in the sorted list (see Figure \ref{fig:schedule}).
\subsection{Extending to Multiple Crossbars}
\label{sec:stride}
Now suppose we have a set $\mathcal{C}$ of $L$ programmable crossbars. Our mathematical optimization model becomes
\begin{equation}
    \{B^{\star}_1, \dots, B^{\star}_L\} = \arg \min_{A_{m\times n}\in \mathcal{C}, \hspace{0.5em} B_{m\times n} \in \mathcal{W}}\sum R_{AB}.
\end{equation}
That is, we want to find a collection of matrix sections $\{B^{\star}_1, \dots, B^{\star}_L\}$ that minimizes reprogramming of $L$ crossbars. 

We propose the stride $L$ and the stride $1$ scheduling.

\textbf{Stride $\bm{L}$ Scheduling}.
We select the $L$ first crossbars in the sorted list, At each reprogramming time, we program the crossbar $i+L$ to the crossbar $i$.

\textbf{Stride 1 Scheduling}.
We select $L$ evenly spaced crossbars in the sorted list, At each reprogramming time, we program the crossbar $i+1$ to the crossbar $i$.

\begin{figure}
    \centering
    \includegraphics[width = \columnwidth]{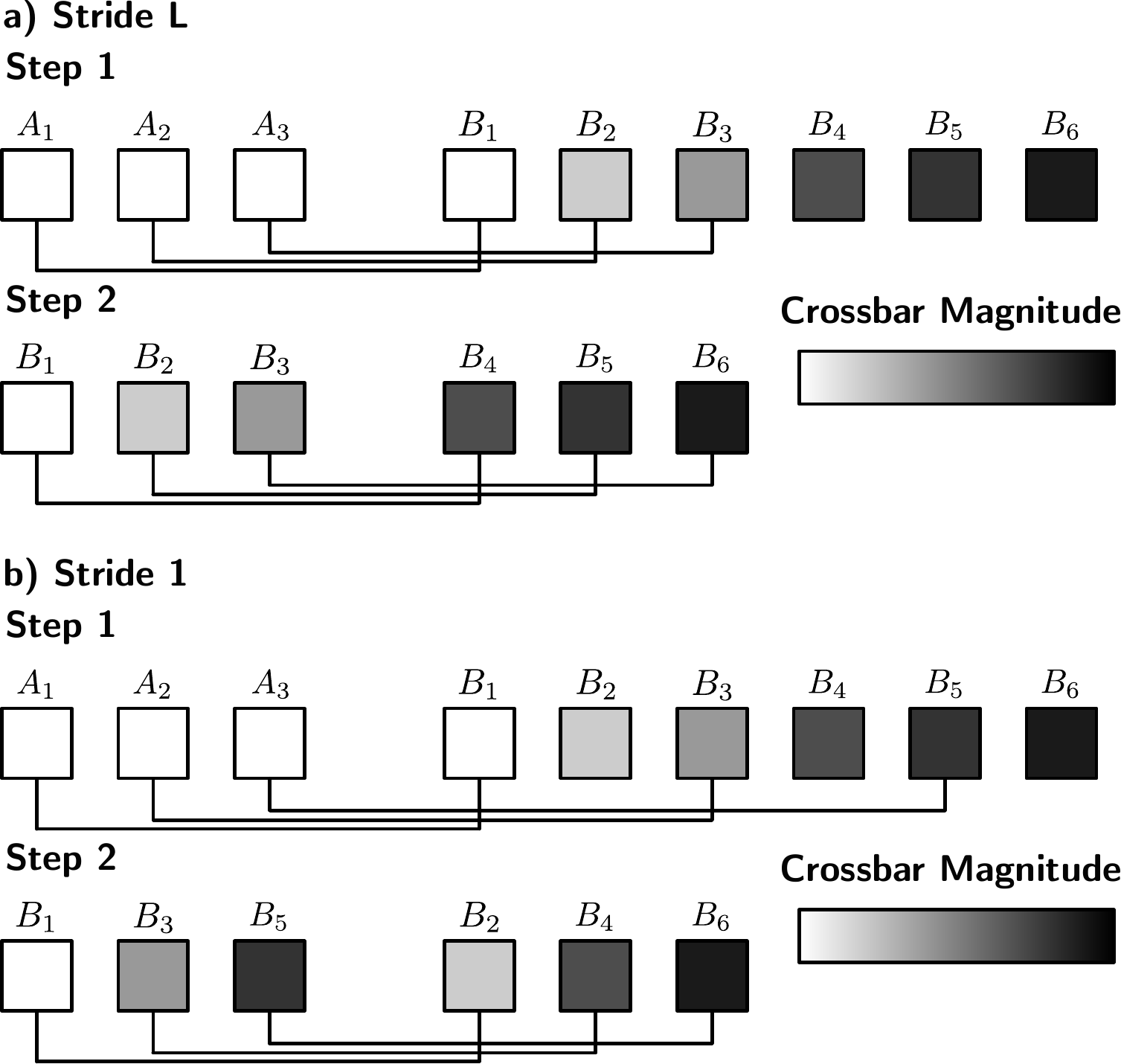}
    \caption{Different scheduling methods for programming multiple crossbars a) stride $L$ and b) stride 1.}
    \label{fig:strides}
\end{figure}

The stride $L$ efficiency depends on the skip length $L$ and the similarity of sections in the sorted list. Although stride 1 initially incurs higher costs by programming the first $L$ crossbars, it only skips one crossbar at each step, making it advantageous over multiple reprogramming tasks, as demonstrated in Figure \ref{fig:strides}. In CIM, write operations for multiplication are over 150x more frequent than in conventional systems, which significantly strains device endurance; for instance, a 1024x1024 resistive RAM crossbar performing 32-bit multiplication reaches failure in about five minutes \cite{endurance}. Additionally, limited physical space in crossbars prevents storage of entire DNN weight matrices, necessitating frequent reprogramming to load different sections. Under these conditions, stride 1 is anticipated to reduce reprogramming effectively as we will see in Section \ref{sec:exp}.


\subsection{Multiple Crossbar Programming Thread Imbalance}
\label{sec:thread}
If we program $L$ crossbars in parallel, we expect to achieve speedup of $L$ times. Realistically, depending on how we balance the work of each thread, it might result in significant delay. For instance, if we group crossbars of small and large reprogramming costs within the same thread, the programming time will be bottlenecked by the largest reprogramming cost.

\begin{figure}
    \centering
    \includegraphics[width = \columnwidth]{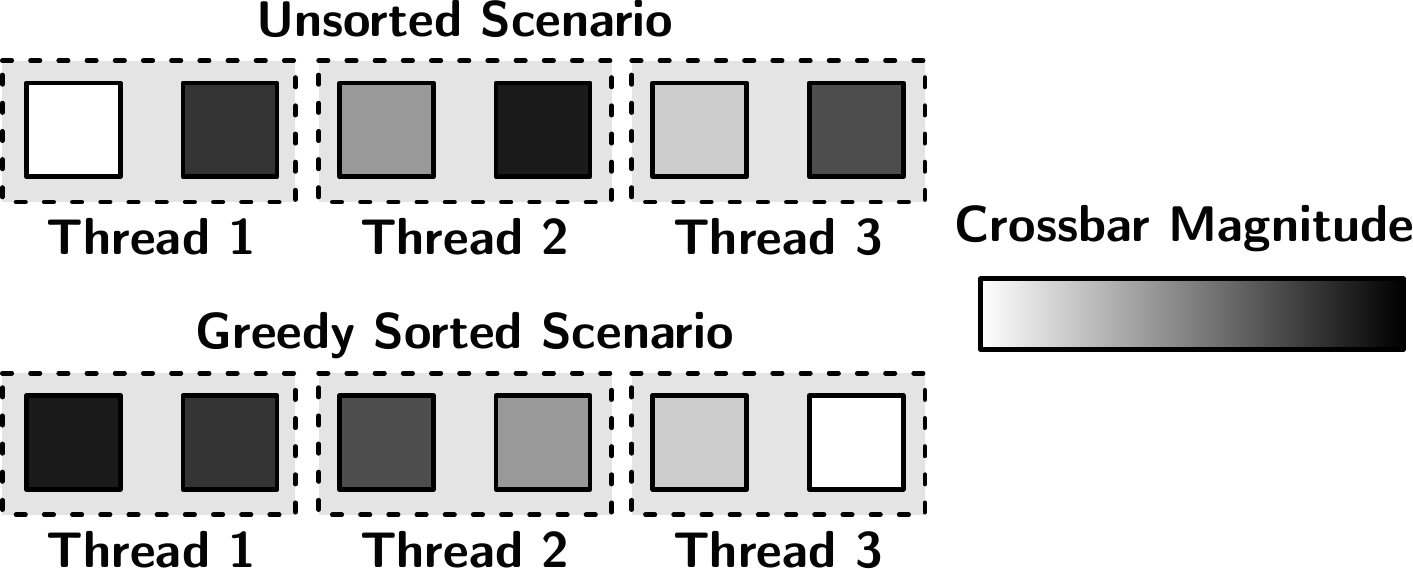}
    \caption{We group crossbars of similar cost on each thread to boost speedup when parallel programming multiple crossbars.}
    \label{fig:greedy}
\end{figure}

To optimize parallel programming of multiple crossbars, we propose a greedy approach based on SWS. We group crossbars of similar reprogramming costs on each thread (see Figure \ref{fig:greedy}).

\section{Bit Stucking-Based Reprogramming}
\label{sec:approx}
Now we leverage bit-level weight distribution to further reduce reprogramming with minimal DNN accuracy loss.

Consider a random sample of numbers in bitline representation. The probability that a number has a digit one on its least significant bit (LSB) tends to 50\% as we increase the bitwidth: the distribution of values in the LSB is approximately uniform.

In bit-sliced crossbars, weights are mapped in bitline form on each row. This way, the LSB is the lowest-order column memristor. Since each column is a power-of-two multiplier, the distance between the column and the input dictates its importance on the weight magnitude. That is, the lowest-order column represents the smallest multiplier.

Unstructured pruning is a compressing technique that sets weights below a certain threshold to zero \cite{liang2021pruning}. For CIM crossbars, a similar concept is applied stucking column outputs. In this case, weights are not set to zero; instead, we introduce errors. Notably, altering the lowest-order column impacts all crossbar weights, thus, harsher than unstructured pruning.

Depending on the number of columns, pruning the lowest-order column might have a severe impact on the final accuracy. Noteworthy, low-order columns tend to have more transitional memristors due to their uniform distribution. We provide a DNN accuracy analysis when neglecting some transitional memristors in the last column to understand the reprogramming tradeoff.

\section{Experiments}
\label{sec:exp}
We assess performance using speedup (ratio of memristors that needed to switch states) and model accuracy, benchmarking against the state-of-the-art unsorted scenarios from CASCADE \cite{chou2019} and ISAAC \cite{shafiee2016}. Crossbar computations were simulated in PyTorch on ImageNet-1K \cite{deng2009} on all model layers (ResNets and VGGs from PyTorch, and ViTs and DeITs from \texttt{timm}), trained in 32-bit floating point precision. Simulations utilized 128x10 crossbars unless specified otherwise.



\subsection{Sorted Weight Sectioning on a Single Crossbar}
\label{sec:exp_parallel}
\begin{figure}
    \centering
    \includegraphics[width=\columnwidth]{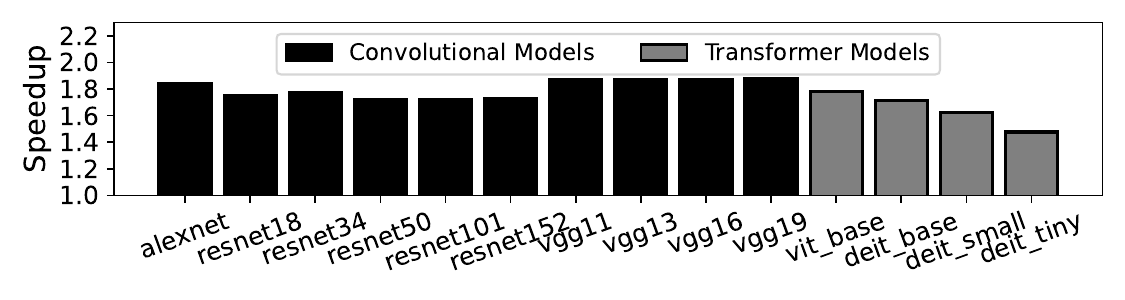}
    \caption{Speedup of SWS for a single 128x16 crossbar.}
    \label{fig:ss_single}
\end{figure}

The DNN bell-shaped distribution \cite{han2015, fang2020, horton2022, tambe2020} results in many crossbars with low-magnitude weights, which require fewer transitional states. The gradual transition from small to large-magnitude crossbars is essential to understanding how SWS optimizes reprogramming. In Figure \ref{fig:ss_single}, DeIT-Tiny, with its sharp weight distribution, achieved the lowest speedup (1.47x), while VGGs, with smoother distributions, saw higher improvements, reaching 1.87x on VGG16. Notably, SWS enhanced programming speed for all models.


\subsection{Sorted Weight Sectioning on Multiple Crossbars}
\label{sec:exp_multiple}
\begin{figure}
    \centering
    \includegraphics[width =\columnwidth]{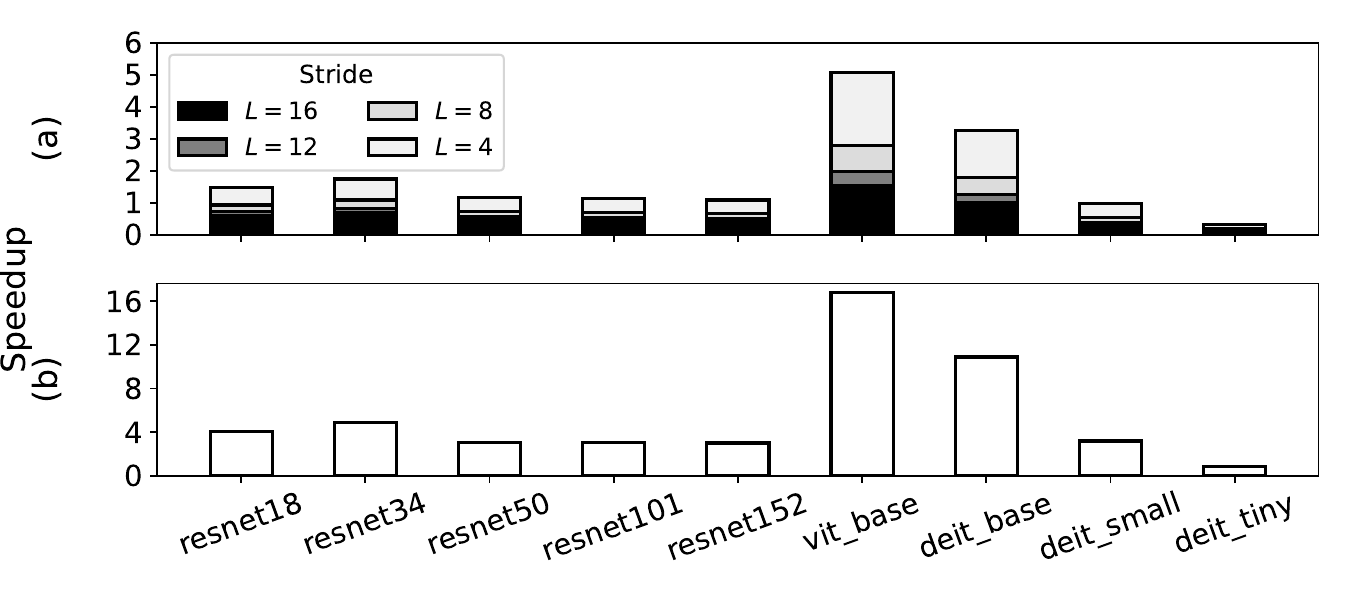}
    \caption{Speedup of parallel programming multiple crossbars with (a) stride $L$ and (b) stride 1 scheduling methods.}
    \label{fig:a}   
\end{figure}

We compare the average speedup per crossbar in two stride methods with 16 reprogrammable crossbars, we observed that increasing the stride reduced SWS speedup, as shown in Figure \ref{fig:a}. This decline in speedup occurs because larger strides require skipping multiple ($L$) crossbars per reprogramming task. As expected, stride 1 yielded the best results; for example, in ViT-Base, stride 1 produced a speedup 3x greater than stride $L=4$, suggesting that the initial cost of programming farther crossbars becomes negligible over time.

\begin{figure}
    \centering
    \includegraphics[width = \columnwidth]{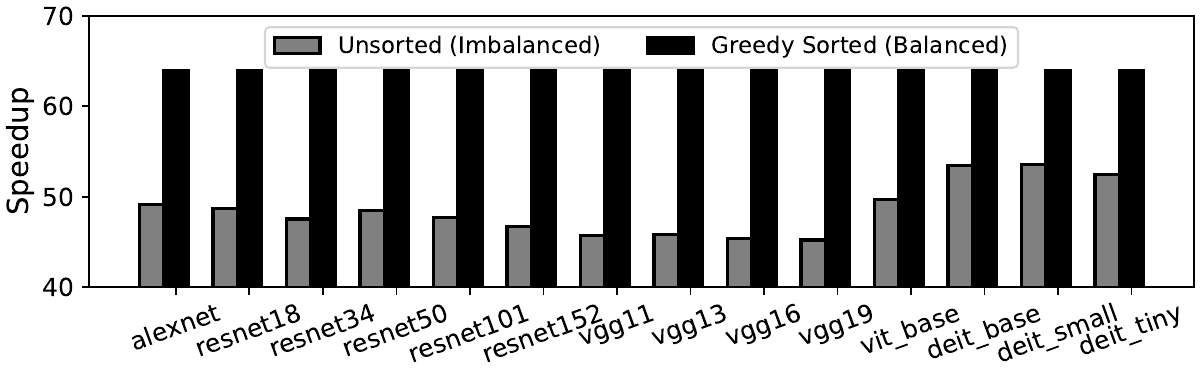}
    \caption{Speedup of greedy approach with 64 crossbar threads.}
    \label{fig:speedup64}
\end{figure}

Now, we analyze the parallel programming of multiple crossbars. The unsorted result observed in Figure \ref{fig:speedup64} is bottlenecked by the slowest crossbars in each thread. We note that VGG models suffered more with thread imbalance due to the disparity between crossbars on each thread. 
With the greedy method, we obtained a result very close to the ideal of 64x speedup.

\subsection{Bit Stucking-Based Reprogramming Analysis}
\label{sec:exp_approx}
\begin{figure}
    \centering
    \includegraphics[width = \columnwidth]{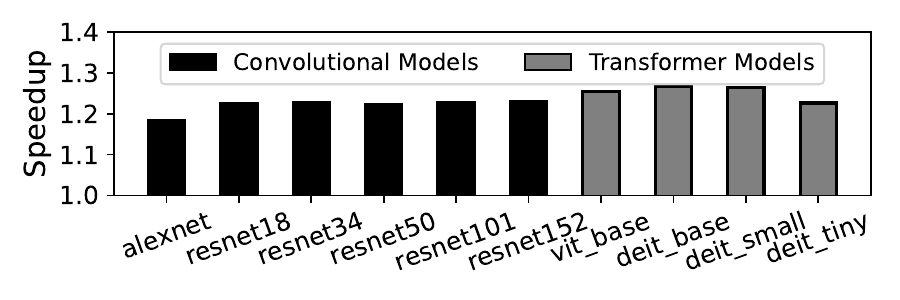}
    \caption{Speedup with $p=0.5$ over $p=1$.}
    \label{fig:p05}
\end{figure}

\begin{figure}
    \centering
    \includegraphics[width =\columnwidth]{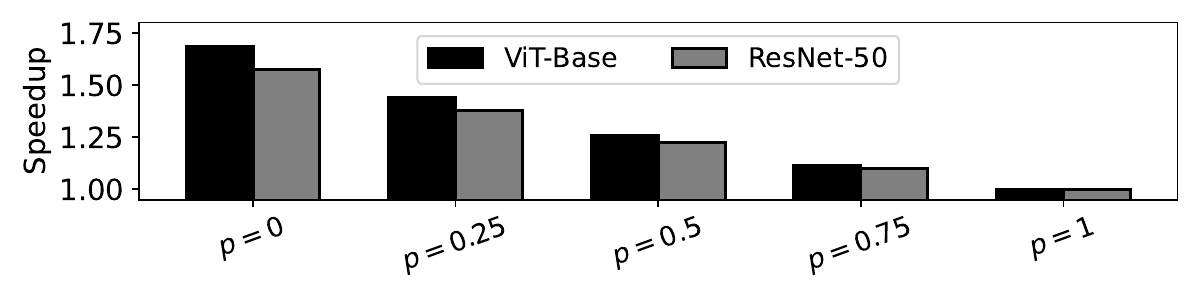}
    \caption{Speedup sweeping $p$ for ViT-Base and ResNet-50.}
    \label{fig:sweepperc}   
\end{figure}

The bit stucking-based reprogramming experiments, shown in Figure \ref{fig:p05}, compare $p=1$ (full reprogramming of necessary memristors in the lowest-order column) with $p=0.5$ (only half are reprogrammed). Speedups ranged from 19\% for AlexNet to 27\% for DeIT-Base, with minimal accuracy loss (less than 1\%). A further sweep of $p$ values for ViT-Base and ResNet-50 in Figure \ref{fig:sweepperc} demonstrates that reducing $p$ down to 0—essentially stucking the last column—maintains accuracy within 1\%. Overall, tuning $p$ between 0 and 1 effectively balances speedup and accuracy.

\begin{figure}
    \centering
    \includegraphics[width =\columnwidth]{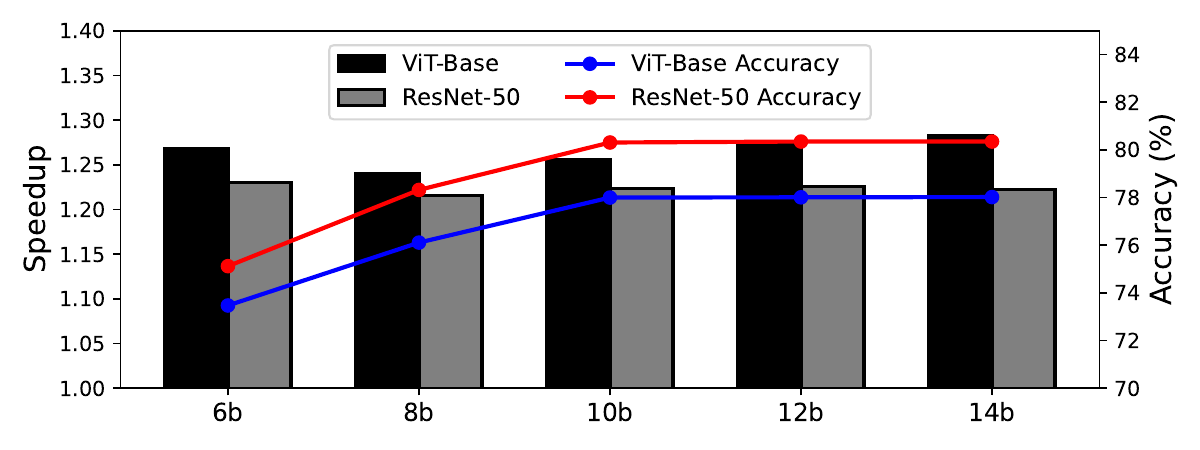}
    \caption{Speedup and accuracy of bit stucking-based reprogramming sweeping number of crossbar columns for ViT-Base and ResNet-50 with probability $p=1$ over $p=0.5$.}
    \label{fig:sweepcols}   
\end{figure}

Finally, we fix $p=0.5$ and sweep the number of columns (see Figure \ref{fig:sweepcols}). Sweeping columns provided almost constant speedup, while the accuracy reaches a plateau in 10 columns (78.00\% and 80.31\% in ViT-Base and ResNet-50, respectively).

The accuracy drop before stabilization occurs because reducing crossbar columns lowers weight bitwidths, and stucking low-order columns significantly affects the bitline representation for lower bitwidths. Nonetheless, bit stucking-based reprogramming using SWS at stride 1 and $p=0.5$ achieved substantial speedups—3.7x for ResNet-50 and 21x for ViT-Base—while maintaining less than 1\% accuracy drop.


\section{Conclusion}
We showed that sorting pretrained DNN weights for bit-sliced CIM crossbars and leveraging memristor distribution in low-order columns significantly save reprogramming time by minimizing transitional resistance states. The method's effectiveness depends on how parameters are distributed in crossbars: bit-level sparsity, models with similar matrices, and large-sized crossbars provide better results.

We validate these results on ImageNet-1K dataset. We achieved substantial speedups—3.7x for ResNet-50 and 21x for ViT-Base—while maintaining less than 1\% accuracy drop. This work suggests a new research direction on efficient crossbar reprogramming based on weight allocation.

\bibliographystyle{ieeetr}
\bibliography{refs}
\end{document}